# High Electron Mobility in Epitaxial Graphene on 4H-SiC(0001) via post-growth annealing under hydrogen

E. Pallecchi[1], F. Lafont[2], V. Cavaliere[1], F. Schopfer[2], D. Mailly[1], W. Poirier[2] and A. Ouerghi[1*]

[1] Laboratoire de Photonique et de Nanostructures (CNRS - LPN), Route de Nozay, 91460 Marcoussis, France
[2] Laboratoire National de Métrologie et d'Essais, 29 Avenue Roger Hennequin, 78197 Trappes, France

**ABSTRACT**

We investigate the magneto-transport properties of epitaxial graphene single-layer on 4H-SiC(0001), grown by atmospheric pressure graphitization in Ar, followed by $H_2$ intercalation. We directly demonstrate the importance of saturating the Si dangling bonds at the graphene/SiC(0001) interface to achieve high carrier mobility. Upon successful Si dangling bonds elimination, carrier mobility increases from 3 000 $cm^2V^{-1}s^{-1}$ to > 11 000 $cm^2V^{-1}s^{-1}$ at 0.3 K. Additionally, graphene electron concentration tends to decrease from a few $10^{12}$ $cm^{-2}$ to less than $10^{12}$ $cm^{-2}$. For a typical large (30 × 280 $\mu m^2$) Hall bar, we report the observation of the integer quantum Hall states at 0.3 K with well developed transversal resistance plateaus at Landau level fillings factors of $\nu$ = 2, 6, 10, 14… 42 and Shubnikov de Haas oscillation of the longitudinal resistivity observed from about 1 T. In such a device, the Hall state quantization at $\nu=2$, at 19 T and 0.3 K, can be very robust: the dissipation in electronic transport can stay very low, with the longitudinal resistivity lower than 5 m$\Omega$, for measurement currents as high as 250 $\mu$A. This is very promising in the view of an application in metrology.

**KEYWORDS:** Epitaxial graphene – Hydrogenation – magneto-transport properties – Electronic properties – Quantum Hall effect

* Correspondence to: abdelkarim.ouerghi@lpn.cnrs.fr

Graphene is a monoatomic layer of carbon atoms that possesses unique electronic properties. It is a zero band gap semiconductor, therefore ambipolar; it has extremely high carrier mobility at room temperature and charge carriers behave as massless Dirac fermions[1]. Over the last years, the development of techniques for synthesizing wafer-scale graphene has strongly accelerated and dramatically expanded the potential for applications of this new material to energy storage, supercapacitor, metrology, spintronics, electronics, and optoelectronics applications[2]. Among these techniques, the graphitization of silicon carbide wafer by controlled sublimation of silicon atoms is promising for electronic applications since it allows the direct growth of large domain, high mobility graphene on a high quality insulating substrate[3]. Moreover, the integration to conventional device technology and molecular electronics appears feasible.

The quality and number of layers in epitaxial graphene on SiC depend on the growth conditions (pressure, temperature), SiC polytype, face and orientation. Atmospheric pressure graphitization of a SiC(0001) substrate can be used to produce large terraces fully covered by a single layer graphene[4,5]. For Si terminated SiC substrates the first carbon layer which is formed, the so-called interface or buffer layer, is insulating because one third of its carbon atoms are covalently bound to the substrate. The epitaxial graphene refers so to the carbon layer formed on top of this interfacial layer, and features linear dispersion typical of isolated single layer graphene[6]. However, a high carrier concentration is generally observed because of the charge transfer from the interface density of states associated to the buffer layer and the silicon dangling bonds at the SiC substrate surface[7]. A route to produce less doped graphene, the so-called quasi-free standing epitaxial graphene (QFSEG), is to decouple the interface layer from the substrate by hydrogen intercalation[6]. At the right temperature and pressure conditions, the H atoms can break the Si-C bonds that link the interface layer to the substrate transforming it in a graphene layer. Moreover, hydrogen is able to saturate the Si dangling bonds[8,9]. The consequence of these transformations is a strong suppression of the charge transfer. Although this approach works, mobility in QFSEG is still lower than expected, most likely because of the rather large amount of defects in the interface layer. *Ab initio* calculations have shown that the buffer layer is a warped carbon layer with periodic inclusions of pentagon-hexagon-heptagon (H5, 6, 7) complexes, covalently bond to the Si atoms of the substrate[10]. Alternatively, oxygen adsorption has already be used to reduce the intrinsic carrier concentration in epitaxial graphene[11]. On the other hand, is it not clear whether oxygen adsorption introduces additional disorder into the graphene layer.

Here we propose post-growth gentle $H_2$ annealing as a route to obtain high mobility epitaxial graphene. In this case hydrogen can saturate the silicon dangling bonds without a real decoupling of the interface[11]_ In these conditions the charge transfer to the graphene layer is reduced and, as a consequence, we can expect a decrease of the carrier concentration and an increase of the carrier mobility. Graphene Hall bars were measured at low temperature, from their magnetoresistance we extract mobility and carrier concentrations. When comparing intrinsic and gently hydrogenated graphene, beyond a slight decrease of the carrier concentration below $1\times10^{12}$ cm$^{-2}$, we find, more importantly, a clear increase of the mobility by about a factor of 2 to 5. The treatment is robust against nanofabrication and thermal cycling, from room temperature down to cryogenic temperature. For hydrogen treated graphene, we find motilities as high as 11 300 cm$^2$V$^{-1}$s$^{-1}$ at 0.3 K and the quantum Hall effect confirms the high quality of the graphene obtained, making the developed growth process a promising route for demanding applications like metrology[12].

The graphene used in this study is obtained by annealing 4H-SiC(0001) at 1550 °C in 800 mbar argon for 10 min[13] [see Methods]. We produced two graphene samples, labeled A and B, in two subsequent runs using the same growth conditions. In order to study the effect of saturation of the Si dangling bonds at the interface layer, the sample B was cut into two parts, one of which was exposed to hydrogen. The hydrogenation was carried out using 100% $H_2$ at 820 °C for 10 mn, in the following we will refer to this sample as to $BH_2$. The goal here is to saturate the Si dangling bonds from the substrate without decoupling the interface layer from the substrate (figure 1(a)). We cannot exclude, a priori, the possibility to break the covalent bonds Si-C in such conditions, but we give experimental evidence a posteriori that such complete decoupling does not happen[14]. We probed the electronic properties of the samples using photoemission spectroscopy (XPS). The C $1s$ XPS spectrum of the sample B and $BH_2$, collected for a photon energy hν = 340 eV, is shown in figure 1(b). The C $1s$ spectrum (sample B) showed three components at 283.7, 284.6, and 285.4 eV in binding energy[4,13]. These components correspond to the SiC bulk (noted SiC), the graphene layer (noted G), and the interface layer (noted IL), respectively. For the $BH_2$ sample, the C $1s$ spectrum shows the same three components (G, IL and SiC). However, the SiC bulk component in the C$1s$ core level spectra shifts 0.2 eV towards lower binding energy. This variation in the band bending in the sample $BH_2$, confirms the partial saturation of Si dangling bond[14] and reveals that the graphene/SiC interfacial region is modified after hydrogenation.

Assuming the graphene-SiC sample can be modeled as a semi-infinite SiC substrate with a uniform graphene overlayer, the thickness of the graphitic film can be calculated from the ratio between the intensity of the G and SiC components of the XPS spectrum. This ratio fits well with an exponential decay of roughly single-layer of graphene covering for the two samples. Additionally, we used Raman spectroscopy to study the effect of hydrogen adsorption. Figure 1(c) shows a typical Raman spectrum of the single layer graphene and single layer graphene with and without hydrogenation (sample B and $BH_2$). Several intense peaks are observed between 1200 and 1800 cm$^{-1}$; they correspond to second-order Raman bands originated from the SiC substrate[15,16]. Graphene contributions are also observed, they are identified by three main structures: i) the D band at 1350 cm$^{-1}$, ii) the G band (symmetric $E_{2g}$ phonon mode) at 1582 cm$^{-1}$ and iii) the 2D band at 2696 cm$^{-1}$. In the two spectra (B and $BH_2$) the D band which corresponds to the disorder is weak in comparison with G and 2D band (double resonant electron-phonon process). The low intensity of this peak shows that there is only a small amount of defects/disorder in the graphene structure. This is an indication for the high quality of the produced epitaxial graphene before and after hydrogenation.

Electron transport in graphene is sensitive to the amount and type of disorder, and when combined with structural and electronic property analysis techniques, it allows in-depth probing of the graphene properties. For this purpose, [see Methods] large 30-μm wide multi-terminal Hall bars were fabricated from hydrogen exposed graphene sample (BH2) and from intrinsic graphene samples (A and B); the latter to be used as reference to evaluate the effect of the hydrogen exposure. Hall bars were designed with the drain-source channel oriented along the step edges of graphene and rotated by 45 and 90 degrees in order to test any anisotropy of charge transport in epitaxial graphene related to the step-structured character of the SiC surface. The graphene terraces are not seen under normal reflection optical microscopy, and while is possible to use AFM this is time consuming, especially to obtain information about large samples as in our case. We found that the graphene terraces can be clearly detected in an optical microscope by using polarized light[17]. This allowed an easy production of Hall bars at well defined angles with respect to the direction of the substrate step edges, as it can be seen in [Figure 2(a) and 2(b)]. The AFM investigation on the graphene produced after the device fabrication shows that residue of optical resists are left on the sample. We point out that the surface is highly uniform at large scale with atomically flat, 2 – 4 μm-wide terraces (figure 2(c)). On defect-free areas of the sample, the terraces typically extend undisturbed over 50 μm in length.

We now discuss the results of electronic transport property measurements of a certain number of devices fabricated with the three produced graphene samples (A, B, for intrinsic graphene and BH2 for gently hydrogenated). We first characterized the room temperature Drude resistivity via 4 terminals probe station measurements. The resistivity is 2.3 and 3.6 kΩ in average for devices from B, and BH2 respectively as shown in Figure 2(d). The difference between the average resistivity values for samples from $BH_2$ and B is significant. It cannot be explained by the irreproducibility of the growth process, since the samples are from the same growth process and this rather suggests the impact of gentle hydrogenation on charge transport in graphene. Finally, it appears that hydrogenation does not deteriorate the homogeneity of the transport property since the width of the resistivity value distribution is not worse in the case $BH_2$ than for B. This finite width is predominantly ascribed to contaminations coming from the sample fabrication and/or to inhomogeneity in the graphene growth[18]. In a second step, longitudinal ($\rho_{xx}$) and transverse ($\rho_{xy}$) magnetoresistivity measurements were carried out. Such Hall measurements allow the extractions of the carrier concentration and the carrier mobility separately, whereas it is not possible to disentangle both quantities in a Drude resistivity measurement solely. Moreover, at higher magnetic field, the relativistic quantum Hall effect, the hallmark of massless Dirac fermions, gives a finer insight of the electronic transport properties quality. We present on Figure 3 the results for three typical devices corresponding to sample A, B and $BH_2$ (top to down panel). For the device from sample A, the relative angle between the SiC step edges and the Hall bar channel is 10°, for the devices from B and $BH_2$, it is 45°, the same orientation making the comparison easier. The zero magnetic field peaks in the longitudinal resistivity is due to the weak localization correction to the conductivity, and is well known in disorder metals, in 2DEGs, and graphene[19,20]. In these samples, it is a small correction, on the order of a few hundred ohms, and it is confined at low fields around zero, a first indication of the relatively high mobility of our graphene. For the three devices $\rho_{xx}$ starts to oscillate when the magnetic field is increased. These oscillations, known as Shubnikov de Haas oscillations, appear with the Landau quantization of the density of states. At higher magnetic fields, up to 11 T, the QH effect develops: plateaus are observable in the transverse resistivity $\rho_{xy}$ evolution with the magnetic field in coincidence with the $\rho_{xx}$ minima[21]. We first focus the attention on the low temperature carrier concentration and mobility of the different samples extracted from the low magnetic field magnetoresistivity measurements. The carrier concentration is easily extracted from the slope of $\rho_{xy}$ versus B, in the regime of the classical Hall effect; it is equal to $n=B/(\rho_{xy}.e)$. The mobility μ is then extracted from the zero field Drude resistivity $\sigma_0 = 1/\rho_0 = ne\mu$. To determine the Drude resistivity, we subtracted from the zero field value the weak localization peak. The electron-electron contribution to the conductivity is low and can be neglected because of the large value of the diffusion constant (ref e-e interaction)[19]. The effect of hydrogenation on electronic transport properties can be easily accessed by comparing the two devices fabricated from B and $BH_2$.

The carrier density for the three n-type doped devices ranges from 1 to $2\times10^{12}$ cm$^{-2}$. The typical devices made from A and B present carrier concentrations which are different by a factor of two. That difference could be explained by the slight irreproducibility of the growth process, since these samples are from 2 subsequent growths, and/or by differences in the fabrication process, the preparation or the conditioning of the samples assuming that the carrier concentration would be over all determined by extrinsic contamination. More interestingly, the carrier density is twice lower in the device made from hydrogenated graphene $BH_2$ when compared to the device made from B. The facts that the two devices originate from the same growth process and were prepared in the same conditions make the interpretation of the carrier concentration comparison between B and $BH_2$ devices more clear and less obscured by the effect of external contamination (ad-atoms). Given that carrier density in the samples is relatively high as compared to the charge neutrality point, in a range where the mobility is not expected to depend too much on density, the comparison between the values of the mobility of the three samples is meaningful and straightforward. For intrinsic graphene the mobility values of the two samples A and B are very close, around 2300 $cm^2V^{-1}s^{-1}$, despite the difference in the orientation of the devices with respect to the step edges (figure 3(a) and 3(b)). The mobility of the device made from $BH_2$ exposed to hydrogen is about 5450 $cm^2V^{-1}s^{-1}$, more than twice the mobility of the intrinsic graphene. Consistently with the increase of the graphene mobility, SdH oscillations appear at lower fields for the hydrogenated sample $BH_2$ than for A and B. They become observable at about 4 and 1.5 T for sample A/B and $BH_2$ respectively, once $\mu B \sim 1$, so that mobility of the order of 2500 and 6000 $cm^2V^{-1}s^{-1}$ respectively can be (roughly) estimated for these samples. About the impact of the gentle hydrogenation, it is interesting to remark that the sequence of the transverse resistance $\rho_{xy}$ plateaus observed in the QHE regime at higher magnetic fields is unchanged for samples B and BH2 (figure 3(b) and 3(c)). In both case, the plateaus at Landau level filling factor $\nu=nh/eB$ equal to 6, 10 are clearly observed with values approximately equal to the exactly quantized value $R_K/6$, $R_K/10$, with $R_K$ the von Klitzing constant theoretically equal to $h/e^2$. This sequence is characteristic of single layer graphene[22]. This indicates that hydrogenation does not lead to a complete decoupling of the graphene from the SiC substrate. It does not induce a transition between a system consisting in a single layer graphene lying on the interface to a Bernal bilayer graphene, confirming the XPS results. In this latter case, the plateaus sequence should have been different, with values at $R_K/(4i)$ (with i > 0 an integer)[23,24].

In the following of the manuscript we concentrate on devices produced with hydrogenated graphene sample $BH_2$. The goal of a first series of experiments with these devices was to quantify the degree of anisotropy of the magnetotransport properties with respect to the angle formed between the Hall bars source-drain channel direction and the substrate step edge overall direction. For that purpose, three devices were compared: one device was aligned parallel to the step edges, the second was rotated by an angle of 45° and the third by 90°. The devices are 30 μm wide and the typical width of the SiC terraces after growth is 4-5 μm as shown in Figure 2(c). In figure 4, the longitudinal and transverse resistivities are plotted as a function of the magnetic field for the three devices. The qualitative behaviors are similar: a small weak localization peak around zero magnetic field, well pronounced SdH oscillations, quantum Hall plateaus around $\nu$ = 2, 6, 10, 14. All the $BH_2$ devices show a mobility, ranging from 4 300 to 6 300 $cm^2V^{-1}s^{-1}$ at 1.7 K, that is about two to three times larger than what was found on devices from samples A and B. Our results show that, as expected, the carrier density is not significantly affected by the alignment, since the Hall bar to Hall bar variation is less than 20%, on the same order of the variation between different parts of the same Hall bar. On the other hand, the carrier mobility shows a clear tendency: decreasing at higher Hall bar misalignments, which is consistent with what is already reported in literature for single layer graphene coupled to the SiC substrate via a buffer layer[25,26]. The most striking consequence of the alignment of the Hall bar with the step edges is that it allows a less dissipative edge channel transport, with the longitudinal resistivity at the Hall plateau that decreases down to lower values for Hall bar along the step edges.

Three 30 × 280 µm$^2$ large Hall devices fabricated from BH$_2$ and oriented parallel to the step edges were characterized at low temperature (0.3 K) several times. They show typical electron doping around $1\times10^{12}$ cm$^{-2}$ (with a minimum of $7\times10^{11}$ cm$^{-2}$) and electron mobility up to 11 300 cm$^2$V$^{-1}$s$^{-1}$. These transport properties (*n* and *µ*) are reproducible upon thermal cycling up to room temperature and can be spatially homogeneous within 20 % in a given of these large device. To demonstrate the potential of hydrogenated graphene, more detailed transport properties measurements have been performed in one of these devices that presents high and homogeneous carrier mobility (8000 – 9600 cm$^2$V$^{-1}$s$^{-1}$) for carrier density of $1\times10^{12}$ cm$^{-2}$ (figure 5(a)). At low magnetic field, weak localization corrections have been evaluated by magnetoresistance measurements, allowing the determination of the phase coherence length to 0.6 µm and the intervalley scattering length to 0.6 µm at 0.3 K by fitting with the adequate theoretical model[27]. Contact resistance has been measured at 0.3 K in a 3-terminal configuration in the QHE regime (plateau at ν=2)[12]. The 30 µm wide current contacts present a resistance lower than 10 Ω. The resistance of the smaller voltage contacts ranges from 500 to 1500 Ω. At 0.3 K, the QHE is observed by magnetotransport measurements between -19 T and 19 T. The longitudinal resistivity shows Shubnikov de Haas oscillations, well resolved from roughly 1T with a first minimum corresponding to the onset of the DOS quantization into the Landau levels with filling factor ν = 42 (figure 5(b)). The Hall resistance also clearly exhibits plateaus at $R_K$/14, $R_K$/10, $R_K$/6, $R_K$/2 when increasing the magnetic field, corresponding to ν = 14, 10, 6, 2. The plateau at $R_K$/18 is even barely visible at ± 2.5 T. In view of the application to metrology for the quantum resistance standard, the exactness and accuracy of the Hall resistance was tested on the plateau at ν = 2 at 19 T and 0.3 K with the metrological instrumentation. An agreement with the expected value $R_K$/2 realized in a GaAs/AlGaAs based quantum resistance standard was found within the relative uncertainty of 5 x 10$^{-3}$ which is not sufficient for the targeted application. Since the $R_H$ quantization level is intrinsically related to the non-dissipative character of the transport along the device[28], we have carried out additional precise longitudinal resistivity measurements in order to understand the poor accuracy. $ρ_{xx}$ can be very low (<5 mΩ) up to very high measurement currents of 250 µA, what is very highly promising for the metrological application. The measured $ρ_{xx}$ value on the plateau at ν=2 in this device approaches the reference values of 0.1 mΩ usually measured up to currents of 100 µA in typical 300-µm wide Hall device used as quantum resistance standard with accuracy of 10$^{-9}$. Nevertheless, very high QHE breakdown currents (current at which the $ρ_{xx}$ starts to exponentially deviating from zero) are not observed along the whole hydrogenated graphene sample (Figure 5(c)). Whereas on one given edge of the device, $ρ_{xx}$ value was measured below 5 mΩ, on the opposite edge it was measured equal to 150 Ω at current as low as 10 nA. The very low breakdown current measured in a certain configuration confirms the local high quality of the graphene grown and post-hydrogenated. Homogeneity and improvement of the transport properties will probably be achieved by fine tuning the growth parameters and optimisation of the device process, e.g. by using e-beam lithography which is known to leave less residues on the graphene surface. More generally, the post-growth annealing of graphene under hydrogen, as proposed in this article, is interesting for the metrological application of the graphene QHE in the sense that it would reduce the carrier concentration and increase the carrier mobility, two inescapable prerequisite for implementing the QHE at low magnetic field which is the base of the graphene promise for that application.

At the very end of the measurement run for the transport property characterization of the devices on the $BH_2$ wafer, mild annealing experiments have been carried out under vacuum ($10^{-5}$ mbar) (see methods)[11]. The evolution of the transport properties (electronic density and mobility) of one particular sample with moderate mobility has been measured, both at 300 K and 0.3 K after each step of the annealing process (see Table). The main effect of the annealing up to 450 K is to significantly increase the low temperature electron density from $1.4 \times 10^{12}$ cm$^{-2}$ to $7 \times 10^{12}$ cm$^{-2}$ with a final value in the range of the typically measured values for non-intentionally treated graphene, coupled to the Si-face SiC via a buffer layer, like sample B. Heating at higher temperature on longer time scale would certainly lead to higher electron doping up to the order of $10^{13}$ cm$^{-2}$. A decrease of the mobility is also importantly observed from 4520 cm$^2$V$^{-1}$s$^{-1}$ to 3000 cm$^2$V$^{-1}$s$^{-1}$ at low temperature, approaching the mobility measured for intrinsic graphene samples A and B. With a closer analysis of the data, it appears that the spatial dispersion of the mobility values at low temperature first increases and then decreases. It seems to reach an intrinsic limit. The temperature dependence of the mobility between 300 K and 0.3 K is rather high before annealing, displaying an increase by a factor of about 3 when decreasing the temperature and then supporting the rather good quality of graphene with a limited quantity of defects. Along the annealing process, it tends to decrease, reaching a minimum after the first annealing step at 400 K. It is compatible with a degradation of the graphene quality. For the electron density, we observe a continuous increase with a final value which does no more depend on temperature, as expected for an intrinsic high doping by charge transfer from the graphene/SiC substrate interface. The annealing step at 400 K during 0.5 h has a crucial impact. It can be interpreted as yielding a partial hydrogen desorption at the interface between the graphene layer and the SiC substrate. In this hypothesis, the complete hydrogen removal would be achieved with longer annealing at higher temperatures. The degradation of the charge transport properties after the annealing gives an additional proof of the benefits of the post-growth treatment by hydrogen. After each annealing step, the longitudinal and transversal magneto-resistivities have also been measured up to 19 T at 0.3 K. The Hall resistance plateau sequence characteristic of single graphene is preserved. After the annealing at 450 K, we note the disappearance of the longitudinal magnetoresistivity asymmetry with the magnetic field direction that is observed in one particular measurement configuration with this modest mobility sample before the annealing. Such asymmetry is also observed in other devices of this study (see fig. 3 and 4 for instance) and is commonly reported in SiC epitaxial graphene. The disappearance of the asymmetry with increasing the carrier density above a few $10^{12}$ cm$^{-2}$ attests the hypothesis that such distortion would predominantly result from the spatial inhomogeneity of the carrier concentration, rather than specific scattering by defects. The restoration of the symmetry could also be ascribed to the system becoming more homogeneous: actually, at this stage, mobility values measured along the sample become less scattered (see Table). Finally, by subsequent doping experiments by exposition to air or isopropanol or water or NH3 (vapour or liquid), the low temperature electronic can further change, again decreasing down to around $4 \times 10^{12}$ cm$^{-2}$, while the low temperature mobility staying at 3000 cm$^2$V$^{-1}$s$^{-1}$, never recovers the highest initial value of the hydrogenated sample. This highlights the specific benefit of hydrogenation in the mobility enhancement[29]. The role of external contamination in the carrier density determination is also confirmed.

In summary, we have presented research that directly shows the importance of introducing a hydrogen intercalation step following epitaxial graphene on 4H-SiC(0001). A gentle hydrogenation that does not lead to decoupling of the graphene film, as evidenced by XPS and transport data, allowed us to increase the charge carrier mobility. Hydrogenation results in the saturation of the dangling bonds of the graphene/SiC(0001) interface layer and provides a means for significant improvements in carrier mobility from approximately 3000 $cm^2V^{-1}s^{-1}$ to > 11 000 $cm^2V^{-1}s^{-1}$ at 0.3 K. In hydrogenated graphene, we observe QHE in a large 30 × 280 $\mu m^2$ sample with Hall resistance plateau sequence consistent with single-layer graphene in the presence of a perpendicular magnetic field. These findings corroborated with XPS studies unambiguously identify the single-layer graphene after gentle hydrogenation attesting the absence of complete decoupling of the graphene from the substrate. This post-growth treatment of graphene grown on SiC(0001) by hydrogen is particularly attractive for electronic and metrology device applications, thanks to its high mobility and its demonstrated quality for low-dissipative transport in the QHE regime.

**Methods**

The single layer graphene studied in this paper were produced via a two-step process from substrate of 4H-SiC(0001) (Si-face). Prior to graphitization, the substrate was hydrogen etched (100% H2) at 1550 °C to produce well-ordered atomic terraces of SiC. The terraces cover the whole wafer (about 1 $cm^2$), with a typical step height of about 4-6 nanometers. The SiC sample is heated to 1000 °C in a semi UHV and then further heated to 1525 °C in an Ar atmosphere. This graphitization process results in the growth of an electrically active graphene layer on top of the interface layer. The graphene sample was characterized using Raman spectroscopy, x-ray photoemission spectroscopy and atomic force microscopy. The micro-Raman spectroscopy was performed at room temperature with a Renishaw spectrometer using 532 nm excitation wavelength lasers light focused on the sample by a DMLM Leica microscope with a 50X objective and a power of 5mW with spot size of about 1 $\mu$m. The AFM measurements were realised under ambient conditions, and the images were recorded in noncontact mode. The XPS measurements were performed on the Tempo beamline of the Soleil synchrotron radiation (Saint-Aubin, France). For transport measurements, we fabricate Hall devices by optical lithography, reactive ion etching and metal deposition. The graphene Hall bars were produced using standard optical lithography. We used the image reversal AZ5214 photoresist, palladium/gold (20/100 nm) was employed as a contact material in order to ensure good contact transparency. Our Hall bars are 30 microns wide, the distance between the voltages probes are either 70 or 110 $\mu$m, the number of voltage probes for each bar is either 4 or 8. The largest devices are 30 x 280 $\mu m^2$.

The electronic magnetotransport measurements were carried out at low temperature, either at 1.7 K in Oxford VTI helium 4 refrigerator with magnetic field perpendicular to the sample up to 11 Tesla, or at 0.3 K in an Oxford Heliox helium 3 refrigerator equipped with a Cryogenic Ltd superconducting magnet capable to generate magnetic fields up to 20 Tesla. The final annealing experiment was performed under vacuum ($10^{-5}$ mbar), in the helium 3 refrigerator, heating is achieved by Joule effect, dissipating electrical power in a resistive discrete component attached to the sample holder. The temperature is monitored with a calibrated Pt100 thermometer also mounted on the ceramic sample holder.

**Acknowledgements**

We are grateful to B. Etienne, M. Ridene and A. Lemaitre for fruitful discussions. This work was supported by the French Contracts ANR- 2010-MIGRAQUEL and ANR-2011-SUPERTRAMP, and the RTRA Triangle de la Physique.

Notes: The authors declare no competing financial interest.

**Author contributions:** A.O. grow the graphene sample and conducted the measurements by XPS, E.P. and V.C. characterized it by AFM and Raman spectroscopy. E.P. and V.C. produced the Hall bars and conducted the measurements at 1.7K, F.L., F.S., W.P. carried experiments at 0.3K. E.P., F.L., F.S., W.P. and A.O. analyzed the data, and D.M. supported the experiment. E.P. and F.S. wrote the paper with all authors contributing to the final version. A.O. planned the experiments and supervised the project.

**Figure captions:**

**Figure 1** (Color online)**: Structural Properties of Epitaxial single-layer graphene after hydrogenation.** a) Schematic representation of hydrogenation process: after hydrogen exposure the Si dangling bonds between SiC and interface are saturated by hydrogen. (b) C *1s* XPS spectra for epitaxial graphene after hydrogenation at 820°C at hν=340 eV. XPS measurements were performed at φ = 45° emergency angle with respect to the sample normal. This spectrum shows the presence of the interface layer after hydrogenation. c) Typical Raman spectra of the graphene sample after hydrogenation. Contributions at the G and 2D band are observed, together with a very low signal at the defect band D.

**Figure 2** (Color online)**: Optical microscope and atomic force microscope (AFM) images of the Hall bar device.** a) and b) Optical image of the Hall bar along and perpendicular to the steps edges of the SiC substrate, c) Typical AFM image of the single-layer graphene device after fabrication. , d) Average *Rs* data for different devices measured in single-layer graphene (before and after hydrogenation).

**Figure 3** (Color online)**: Longitudinal (black) and Hall (red) magnetoresistivity at a temperature *T* = 1.6 K measured in a (30 × 280 μm²) Hall bar geometry in graphene samples as-grown and after hydrogenation.** a) Pristine graphene layer (Sample A) b) Pristine graphene Layer (Sample B) c) Graphene layer after hydrogenation (Sample BH$_2$)

**Figure 4** (Color online)**: Anisotropy of the Longitudinal (black) and Hall (red) magnetoresistivity at a temperature *T* = 1.6 K of the graphene sample after hydrogenation**. a) Hall bar perpendicular to the step edges, b) Hall bar rotated by an angle of 45° at the step edges c) Hall bar along the step edges. The devices are 30 μm wide and 280 μm long.

**Figure 5** (Color online)**: a)** Longitudinal (black) and Hall (red) magnetoresistivity in epitaxial single-layer graphene on 4H-SiC(0001) after hydrogenation (sample BH$_2$) at a temperature *T* = 0.3 K for a 30×280 μm² Hall bar aligned along the step edges. The device is n-doped with concentration of 1×10$^{12}$ cm$^{-2}$ and the carrier mobility is about 9000 cm²V$^{-1}$s$^{-1}$, b) The longitudinal resistivity shows Shubnikov de Haas oscillations, well resolved from roughly 1T with a first minimum corresponding with filling factor ν = 42 c) Dependence of the longitudinal resistivity measured in a particular section along the device with the measurement current, at 0.3 K, 19 T, on the Hall plateau at ν=2.

**Table:**
Electron concentration and mobility of a particular Hall device (sample BH$_2$) measured at 300 K and 0.3 K after each step of an annealing process under vacuum (10$^{-5}$ mbar). Temperature and duration of each step are indicated in the first column. Spatial scattering of the electron mobility values measured along the device are also indicated in brackets. The first line corresponds to the electrical properties measured under ambient conditions before the annealing. The second line corresponds to pumping down to 10$^{-5}$ mbar in 20 minutes, at ambient temperature, as required to load the sample into the helium 3 refrigerator used to cool the sample down to 0.3 K. The pumping of the sample alone, at ambient temperature, has little effect leading to a 5 %-increase of carrier mobility and 5%-decrease of carrier concentration.

| | Electron density ($cm^{-2}$) at 300 K / 300 mK | Electron density relative variation 300 K – 300 mK (%) | Electron mobility ($cm^2V^{-1}s^{-1}$) at 300 K / 300 mK | Mobility ratio 300 mK / 300 K |
|---|---|---|---|---|
| 300 K / 0 h | $1.79 \times 10^{12}$ / - | - | $1660 (1\pm1.3\times10^{-2})$ / - | - |
| 300 K / 0.3 h | $1.7 \times 10^{12}$ / $1.4 \times 10^{12}$ | - 18 % | $1745(1\pm2.5\times10^{-2})$ / $4520(1\pm2.4\times10^{-2})$ | + 160 % |
| 400 K / 0.5 h | $4 \times 10^{12}$ / $3.5 \times 10^{12}$ | - 13 % | $1485(1\pm1\times10^{-2})$ / $2900(1\pm\mathbf{13\times10^{-2}})$ | **+ 95 %** |
| 450 K / 15 h | $6.8 \times 10^{12}$ / $6.47 \times 10^{12}$ | - 5 % | $1345(1\pm1\times10^{-2})$ / $3000(1\pm<1\times10^{-2})$ | + 123 % |

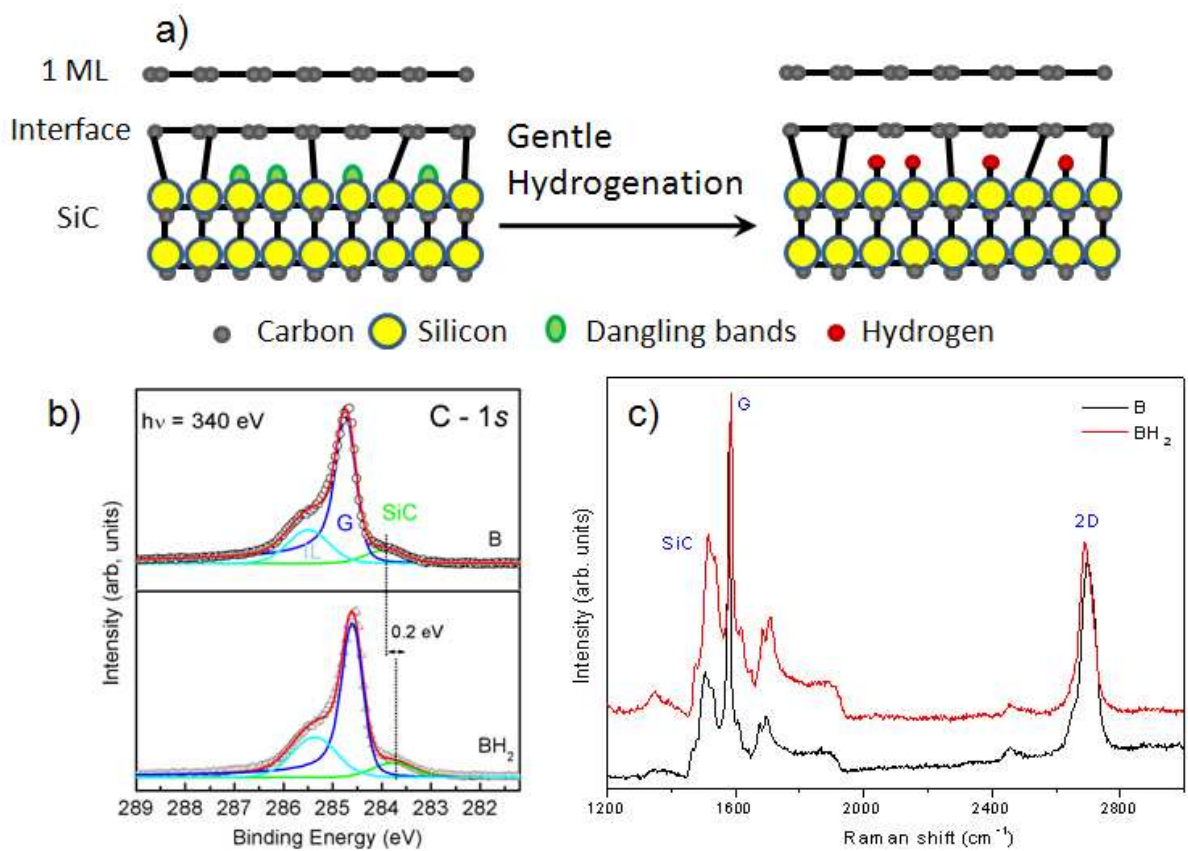

**Figure 1**

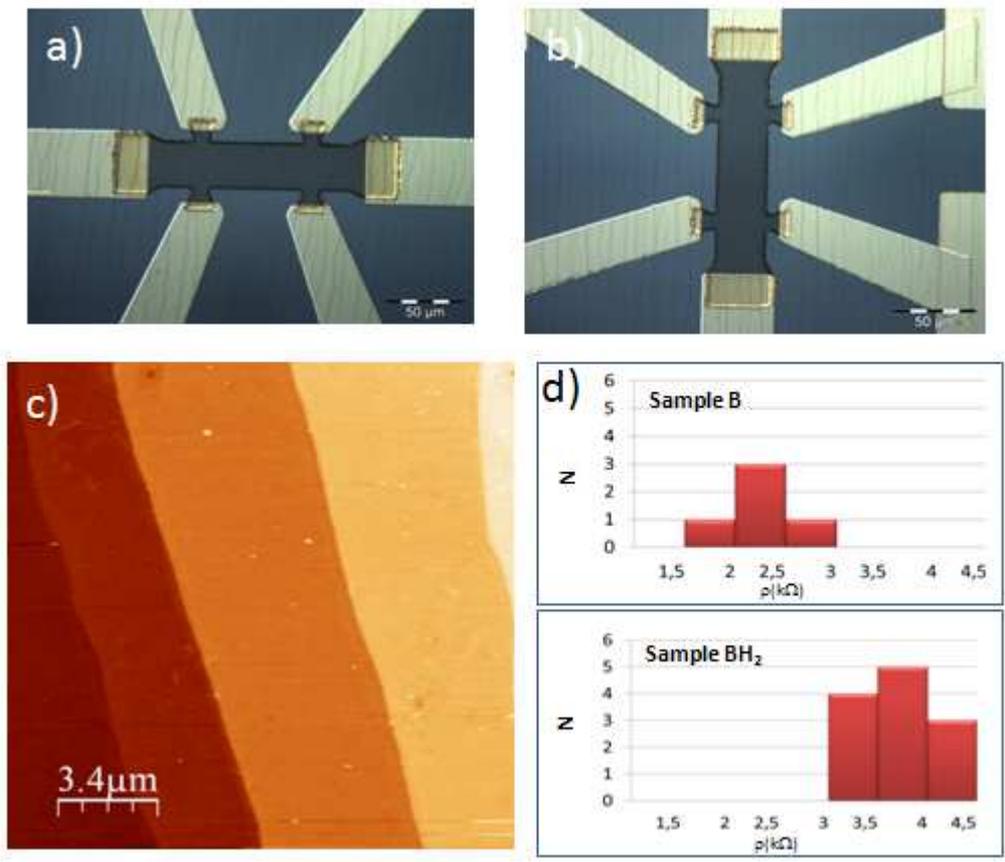

**Figure 2**

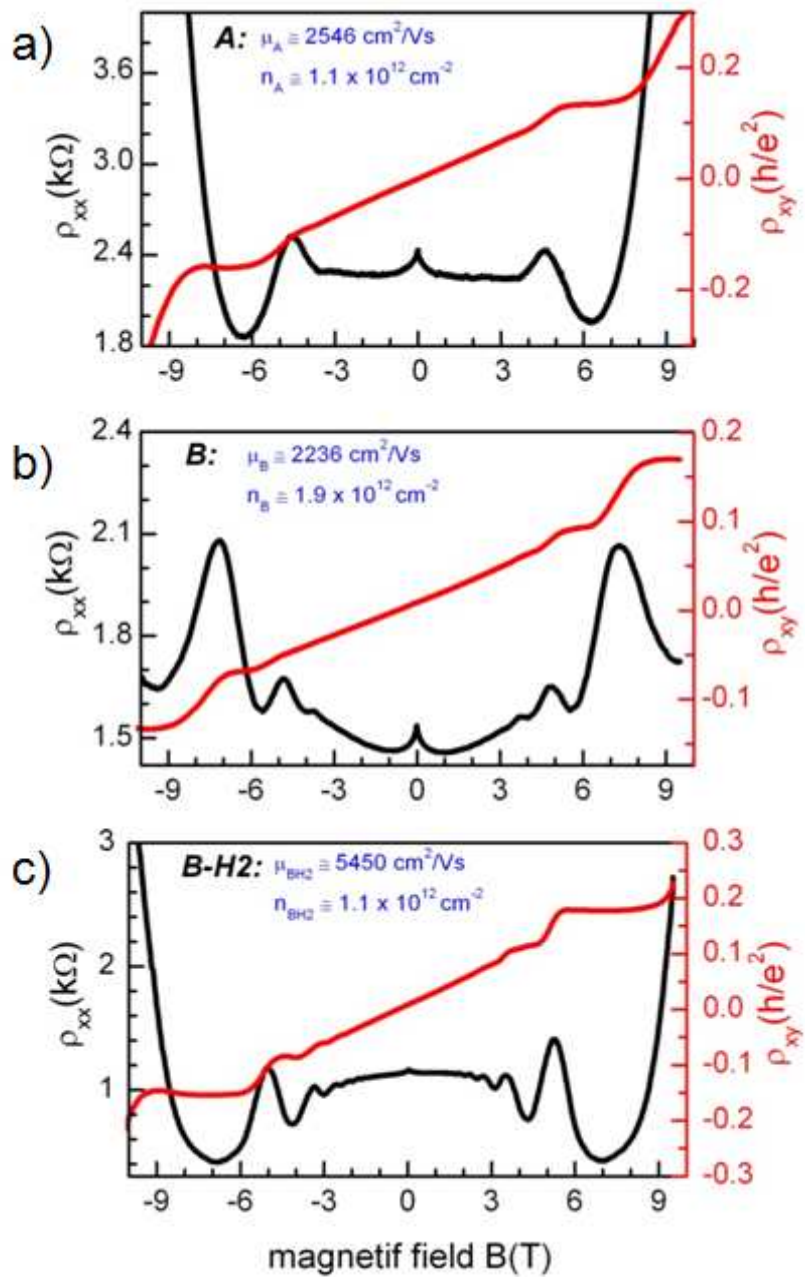

**Figure 3**

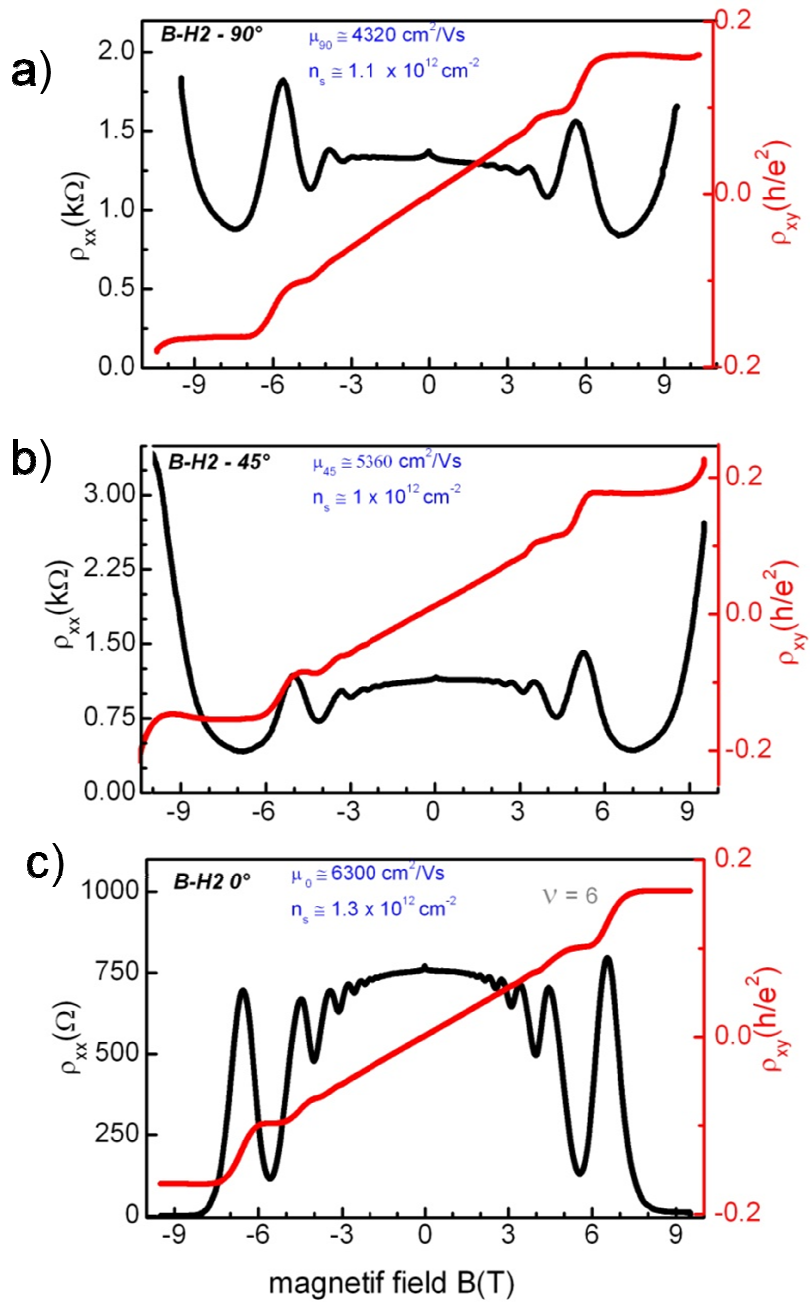

**Figure 4**

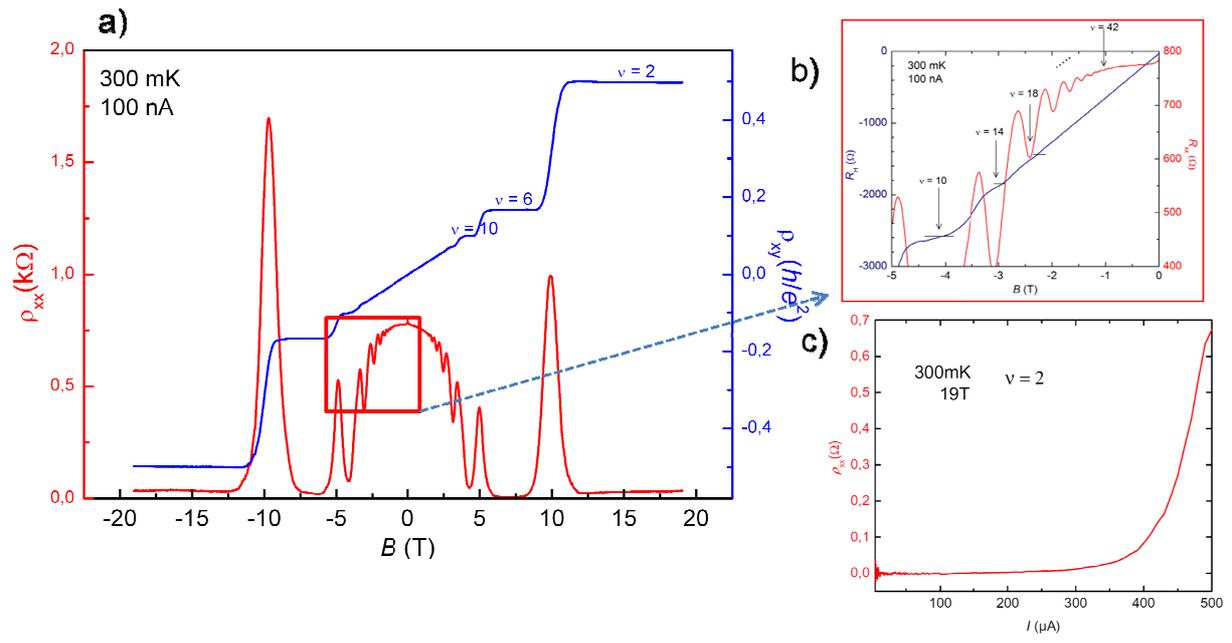

**Figure 5**